\documentclass[aps,prd,twocolumn,10pt]{revtex4-2}

\usepackage{amsmath,amssymb,amsfonts,physics}
\usepackage{graphicx}
\usepackage{hyperref}
\usepackage{bm}
\usepackage{siunitx}
\usepackage{xcolor}
\usepackage{multirow}
\usepackage{booktabs}
\usepackage{lipsum}

\begin{document}

\title{Operator Formalism for Laser-Plasma Wakefield Acceleration}

\author{Mostafa Behtouei, Carlos Salgado Lopez, Giancarlo Gatti}

\affiliation{CLPU, Centro de Láseres Pulsados, Salamanca, Spain}

\begin{abstract}
In this paper, we develop an operator-based framework for laser--plasma wakefield acceleration (LPWA) in capillary discharges, providing a compact and systematic description of the coupled dynamics of laser fields and plasma response. The formalism employs key operators: the transverse modal operator $\hat{K}$, the nonlinear plasma operator $\hat{N}[\Psi]$, the plasma oscillation operator $\hat{\Omega}_p^{\,2}$, and the ponderomotive source operator $\hat{\alpha}$, which together describe mode coupling, plasma oscillations, and nonlinear feedback induced by the ponderomotive force. In the linear regime, the system is characterized by invariant subspaces associated with stable modal structures, while nonlinear interactions break these invariances, leading to mode mixing and complex dynamics. The approach establishes a direct connection between LPWA and Hilbert-space operator theory, including the invariant subspace, providing a formal mathematical interpretation of energy transfer and wakefield formation. Furthermore, the operator formalism integrates with neural operator methods, allowing efficient approximation of $\hat{N}$ and $\hat{\alpha}$ for reduced-order modeling and predictive control. This hybrid physics--AI framework offers a robust foundation for modeling, analysis, and optimization of high-intensity laser--plasma interactions in next-generation accelerator experiments.
\end{abstract}

\maketitle

\section{Introduction}

Plasma wakefield acceleration has rapidly become one of the most compelling directions in modern accelerator physics, driven by the need for more compact and efficient sources of high-energy particles\cite{mangles2004monoenergetic, faure2006controlled, gonsalves2019petawatt, pompili2021energy, pompili2022free, galletti2022stable}. By exploiting the ability of a plasma to sustain extremely large electric fields without material breakdown, these schemes offer accelerating gradients that exceed those of conventional radio-frequency accelerators by several orders of magnitude. This possibility has opened a new path for both fundamental studies and practical applications, ranging from table-top radiation sources to long-term prospects for next-generation high-energy colliders. Within this broader context, wakefields can be driven either by an electron beam-known as plasma wakefield acceleration (PWFA)-or by an intense laser pulse, giving rise to laser-plasma wakefield acceleration (LPWA). While PWFA benefits from the high charge and rigidity of particle beams, laser-driven approaches represent a particularly versatile platform, combining precise control of ultrashort pulses with the unique collective response of a plasma.

In the laser-driven case, laser-plasma wakefield acceleration (LPWA) provides a particularly powerful realization of plasma-based acceleration. When an intense, ultrashort laser pulse travels through a plasma, its ponderomotive force expels electrons from the high-intensity region, thereby generating a large-amplitude trailing plasma wave, or wakefield. This wakefield can sustain accelerating gradients sufficient to drive electrons to relativistic energies within only a few millimeters to centimeters-distances dramatically shorter than those required in conventional radio-frequency accelerators. The behavior of LPWA systems is inherently nonlinear, arising from the intricate coupling between the evolving laser envelope, the resulting plasma-density perturbations, and the complex structure of the induced electromagnetic fields.

Traditional approaches to describing LPWA rely on a combination of Maxwell's equations for the electromagnetic field and fluid or kinetic models for the plasma \cite{esarey2009physics, massimo2020efficient}. In practical scenarios, approximations such as the paraxial approximation are employed, where the laser field is predominantly transverse, and variations along the beam direction are assumed to be slow \cite{esarey2009physics, marceau2013validity}. Within this framework, both the laser and plasma responses are typically decomposed into modes, allowing the system dynamics to be expressed as a set of coupled equations for the modal amplitudes. These equations describe essential phenomena such as linear mode coupling induced by waveguide imperfections, plasma density variations, and nonlinear feedback mediated by the ponderomotive force.

While traditional PDE-based frameworks remain powerful tools for modeling LPWA, they can become increasingly challenging to solve when multiple coupled modes, intricate transverse structures, or strong nonlinearities are present. In these situations, the complexity of the governing equations can obscure the fundamental physical mechanisms-particularly the interactions between different modes and the pathways of energy transfer within the plasma. Consequently, even when numerical simulations are accurate, gaining clear analytical insight can be difficult. To address these challenges, an operator-based formalism offers a powerful alternative. Based on operator theory, this approach describes the laser field and plasma density perturbations classically in terms of mode amplitudes, while treating them mathematically as vectors in an abstract space, with linear and nonlinear interactions described through operators using a formalism analogous to that of quantum mechanics. 

In this formalism, the effects of linear imperfections, such as waveguide geometry deviations or small density fluctuations in the plasma, are represented by a dedicated linear operator that couples different modes and induces phase shifts. Nonlinear interactions due to the plasma response are represented by a nonlinear operator that maps the laser intensity to a self-consistent modification of the mode amplitudes. The intrinsic oscillatory behavior of the plasma is encoded through a plasma oscillation operator, while the ponderomotive force driving the plasma density perturbations is represented by a dedicated source operator. Together, these operators give a clear and compact way to describe the full laser-plasma system, representing the key physics while making the interactions between different modes easier to understand.

The operator-based approach offers several clear advantages for studying LPWA. By focusing on the dynamics of key modes, it provides an efficient framework for describing highly nonlinear laser–plasma interactions and complex couplings between fields and plasma. Linear imperfections, such as waveguide deviations or small density fluctuations, are represented by dedicated operators, while nonlinear plasma feedback, intrinsic plasma oscillations, and the ponderomotive driving force are each encoded through separate operators. This mode-centric formalism makes the pathways of energy exchange explicit, reveals resonances and mode-mixing effects, and dramatically reduces computational cost by replacing full three-dimensional simulations with a smaller set of coupled equations for modal amplitudes. Furthermore, it naturally accommodates both transverse and longitudinal fields, multi-mode laser structures, complex guiding geometries, high-intensity or quantum regimes, and even AI-informed optimization strategies, offering a unified methodology for both analytical studies and practical simulations of laser-plasma wakefield acceleration.

Overall, this introduction establishes the motivation for adopting an operator-based approach to LPWA, highlights the limitations of traditional PDE-based methods, and emphasizes the benefits of a compact, mode-centric, and analytically transparent framework. It provides a context for the detailed construction of linear and nonlinear operators, plasma oscillation dynamics, and ponderomotive driving in the subsequent sections, and sets the stage for systematic studies of energy transfer, mode coupling, and nonlinear phenomena in laser-driven plasma accelerators.

In the following sections, we systematically develop the operator-based framework for laser-plasma wakefield acceleration. We begin by introducing the general formulation of operator dynamics for LPWA and then describe how density perturbations and the ponderomotive force are incorporated into this formalism. Next, we detail the construction of linear and nonlinear operators, including the plasma oscillation operator, and demonstrate the equivalence between the operator approach and the traditional Maxwell–plasma PDEs. We then extend the formalism to full-vector representations, apply Bloch-Floquet theory to analyze mode structures, and explore how the operator framework can be integrated with artificial intelligence for advanced LPWA modeling. Finally, we discuss the connection between operator methods and invariant subspace theory before concluding with a summary of key results and perspectives for future research.

\section{Operator-Based Formulation for Laser-Plasma Wakefield Dynamics}

We now introduce the operator-based formalism for describing laser transport and plasma response in a plasma-filled capillary. The electric field envelope $\Psi(\mathbf r,z)$ in a plasma-filled waveguide satisfies the Helmholtz equation:
\begin{equation}
\nabla^2 \Psi + k_0^2 n^2(\mathbf r,z) \Psi = 0,
\end{equation}
where $k_0 = \omega_0/c$ is the vacuum wavenumber and $n(\mathbf r,z)$ is the local refractive index of the plasma-capillary system.  
The Laplacian is decomposed into transverse and longitudinal parts:
\begin{equation}
\nabla^2 = \nabla_\perp^2 + \partial_z^2.
\end{equation}

Assuming guided-mode solutions,
\begin{equation}
\Psi(\mathbf r,z) = \sum_n A_n(z) \psi_n(\mathbf r,z),
\end{equation}
where $\psi_n(\mathbf r,z)$ are transverse mode functions and $A_n(z)$ are their amplitudes. For a weakly varying or uniform waveguide, the modes $\psi_n$ can be taken as $z$-independent, and variations in permittivity $\Delta \epsilon(\mathbf r,z)$ cause linear coupling between modes:
\begin{equation}
i \frac{d A_n}{dz} = \sum_m \kappa_{nm} A_m, \quad
\kappa_{nm} = -\frac{\omega_0^2}{2} \int \psi_n^* \Delta\epsilon \psi_m \, d^2r.
\end{equation}

Starting from the scalar wave equation in a medium with spatially and temporally varying permittivity $\epsilon(\mathbf r,t)$,
\begin{equation}
\nabla^2 E(\mathbf r,z,t) - \frac{1}{c^2} 
\frac{\partial^2}{\partial t^2} \big[\epsilon(\mathbf r,t) E(\mathbf r,z,t) \big] = 0,
\end{equation}
and assuming a narrowband field along $z$, we write
\begin{equation}
E(\mathbf r,z,t) = \Psi(\mathbf r,z,t) e^{i(k_0 z - \omega_0 t)} + \mathrm{c.c.},
\end{equation}
where $\Psi$ is the slowly varying envelope. Applying the paraxial and slowly-varying-envelope approximations yields
\begin{equation}
i \frac{\partial \Psi}{\partial z} = - \frac{1}{2 k_0} \nabla_\perp^2 \Psi - 
\frac{\omega_0^2}{2 k_0 c^2} \Delta \epsilon(\mathbf r,t) \Psi,
\end{equation}
with $\Delta \epsilon(\mathbf r,t)=\epsilon(\mathbf r,t)-\epsilon_0$ including both capillary imperfections and plasma contributions.  
In a cold plasma \cite{bliokh2017wakefield,aria2008wakefield},
\begin{equation}
\Delta \epsilon_{\rm plasma} = -\frac{\omega_p^2}{\omega_0^2} 
= - \frac{e^2 \delta n_e}{\varepsilon_0 m_e \omega_0^2},
\end{equation}
where $\delta n_e$ is the perturbation in electron density induced by the laser’s ponderomotive force.

To express the coupled laser--plasma dynamics in a compact and basis-independent form, we expand both the laser field envelope and the plasma density perturbation in an orthonormal set of transverse eigenmodes. The laser field is written as
\begin{equation}\label{laser-field}
\ket{\Psi} = \sum_{n} A_n(z,t) \ket{\psi_n},
\end{equation}
and the plasma density perturbation is decomposed as
\begin{equation}\label{plasma-field}
\ket{\delta n} = \sum_{m} \delta n_m(z,t) \ket{\phi_m}.
\end{equation}
The basis functions satisfy the orthonormality relations
\begin{equation}\label{plasma}
\langle \psi_n | \psi_m \rangle = \delta_{nm}, \qquad
\langle \phi_n | \phi_m \rangle = \delta_{nm},
\end{equation}
ensuring that the mode amplitudes form a complete and independent representation of the fields.
With this expansion, all physical effects, linear imperfections, nonlinear coupling, and plasma response can be expressed in a compact operator form. The plasma evolution is governed by
\begin{equation}\label{laser}
\ddot{\ket{\delta n}} + \hat{\Omega}_p^{\,2} \ket{\delta n}
= - \hat{\alpha} \ket{\Psi}^2,
\end{equation}
which describes the driven plasma oscillations under the ponderomotive force.  
The laser evolution obeys
\begin{equation}
i\frac{\partial \ket{\Psi}}{\partial z}
= \hat{K} \ket{\Psi}
- \frac{\omega_0^2}{2 k_0 c^2} \, \hat{N}[\Psi] \ket{\Psi},
\end{equation}
where the operator $\hat{K}$ determines the transverse
modal (including diffraction), $\hat{\Omega}_p^{\,2}$ defines the spectrum of plasma oscillation eigenmodes, $\hat{\alpha}$ measures the ponderomotive coupling from the laser to the plasma, and $\hat{N}[\Psi]$ represents the nonlinear refractive response of the plasma acting back on the laser.

Together, these elements describe the self-consistent interaction loop:
\[
\text{Laser} \xrightarrow{\hat{\alpha}}
\text{Plasma}
\xrightarrow{\hat{N}}
\text{Laser}.
\]
By expressing the laser and plasma fields in an operator-based mode decomposition, we obtain a compact description of their coupled dynamics, including linear imperfections, nonlinear interactions, and plasma response. In the following section, we will construct the explicit operators that govern this self-consistent evolution.

\section{Construction of Operators in Laser--Plasma Wakefield Acceleration}

In LPWA, the process begins when a laser pulse enters the plasma and excites collective electron motion, which in turn engages the plasma's natural oscillatory modes. These oscillations are described by the operator $\hat \Omega_p^2$, representing the intrinsic frequencies of the plasma modes. To construct the plasma oscillation operator $\hat \Omega_p^2$ explicitly, we start from the eigenmodes $\{\phi_m(\mathbf r)\}$ of the capillary plasma, obtained by solving the linearized plasma equations:
\begin{equation}
\nabla_\perp^2 \phi_m + \frac{\omega_{p,m}^2}{v_s^2} \phi_m = 0.
\end{equation}
These eigenmodes form a complete basis for representing any plasma density perturbation, which can then be expanded as
\begin{equation}
\delta n(\mathbf r,z,t) = \sum_m \delta n_m(z,t) \phi_m(\mathbf r),
\end{equation}
where each mode evolves according to its natural frequency:
\begin{equation}
\ddot{\delta n_m} + \omega_{p,m}^2 \delta n_m = 0.
\end{equation}
Projecting the dynamics onto the modal basis, we define the operator
\begin{equation}
\hat \Omega_p^2 \ket{\delta n} = \sum_m \omega_{p,m}^2 \delta n_m \ket{\phi_m},
\end{equation}
which provides a diagonal, Hermitian representation of the intrinsic plasma oscillations.

The laser excites these plasma modes via the ponderomotive force, which is represented by the source operator $\hat \alpha$. This operator is constructed by projecting the squared laser field intensity onto the plasma modes, and is given by

\begin{equation}
\hat \alpha = \sum_m \alpha_m \ket{\phi_m}\bra{\phi_m}, \qquad
(\hat \alpha \ket{\Psi}^2)_m = \alpha_m \int \phi_m^* |\Psi(\mathbf r)|^2 \, d^2 r,
\end{equation}
where $\ket{\Psi} = \sum_n A_n \ket{\psi_n}$ is the laser field in its transverse modal decomposition. The plasma evolution under the laser drive is then
\begin{equation}
\ddot{\ket{\delta n}} + \hat \Omega_p^2 \ket{\delta n} = - \hat \alpha \ket{\Psi}^2,
\end{equation}
showing the immediate formation of the wake.

As the laser travels, its evolution is further influenced by imperfections in the medium, including capillary walls, density fluctuations, or other inhomogeneities. These effects are described by the ransverse
modal operator $\hat K$. Using a complete set of transverse laser modes $\{\psi_n(\mathbf r)\}$, the spatial dielectric perturbation
\begin{equation}
\Delta \epsilon(\mathbf r) = \epsilon(\mathbf r) - \epsilon_0
\end{equation}
is projected onto the modal basis to yield the coupling matrix
\begin{equation}
\kappa_{nm} = -\frac{\omega_0^2}{2} \int \psi_n^*(\mathbf r) \, \Delta \epsilon(\mathbf r) \, \psi_m(\mathbf r) \, d^2 r,
\end{equation}
from which the operator is assembled as
\begin{equation}
\hat K = \sum_{n,m} \kappa_{nm} \ket{\psi_n}\bra{\psi_m}.
\end{equation}
This operator redistributes laser energy among modes due to linear imperfections and diffraction, providing the first feedback from the medium onto the pulse.

Finally, the nonlinear plasma response to the laser intensity is encoded in the operator $\hat N$. The plasma density perturbation $\tilde n(\mathbf r,z,t)$ induced by the laser is projected onto the laser transverse modes:

\begin{equation}
\begin{aligned}
\mathcal N_{nm}(z,t) 
&= \langle\psi_n|\tilde n|\psi_m \rangle
   = \int \psi_n^*(\mathbf r)\, \tilde n(\mathbf r, z, t)\, \psi_m(\mathbf r)\, d^2 r, \\[6pt]
\hat N 
&= \sum_{n,m} \mathcal N_{nm}(z,t)\, \ket{\psi_n}\bra{\psi_m}.
\end{aligned}
\end{equation}

Acting on the laser field $\ket{\Psi}$, this operator redistributes energy among modes:
\begin{equation}
(\hat N \Psi)_n = \sum_m \mathcal N_{nm} A_m,
\end{equation}
accounting for self-modulation, cross-mode energy transfer, and coherent shaping of the wake.

Through this sequence of constructions, $\hat \Omega_p^2$ representing plasma oscillations, $\hat \alpha$ driving them, $\hat K$ describing linear imperfection effects, and $\hat N$ representing nonlinear plasma feedback, the full LPWA process can be observed from the initial laser entry to the self-consistent evolution of both laser and wakefield, allowing one to follow all interactions throughout the acceleration process.

\section{Ponderomotive Force and the Resulting Nonlinear Density Perturbation}

When an intense laser pulse passes through a plasma, the rapidly oscillating electric field exerts a cycle-averaged ponderomotive force that pushes electrons away from regions of high intensity. This process produces a slowly varying, low-frequency density perturbation \(\tilde n\), which in turn modifies the refractive index and feeds back to the laser evolution. The resulting nonlinear response is at the heart of nonlinear laser–plasma interaction phenomena such as self-focusing, self-modulation, and wakefield generation.

We consider a cold, collisionally damped plasma with equilibrium density \(n_0(\mathbf r)\) and small perturbations in density and velocity:

\begin{equation}
n(\mathbf r,t)=n_0+\tilde n(\mathbf r,t), \qquad \mathbf v(\mathbf r,t)=\tilde{\mathbf v}(\mathbf r,t)
\end{equation}
The high-frequency laser field is represented as

\begin{equation}
\mathbf E(\mathbf r,t)=\Re\{\Psi(\mathbf r,z,t)e^{i(k_0 z-\omega_0 t)}\}
\end{equation}
where \(\Psi\) is the slowly varying envelope. The cycle-averaged ponderomotive potential is \cite{silva1999ponderomotive}

\begin{equation}
U_p(\mathbf r,t)=\frac{e^2}{4m_e\omega_0^2}|\mathbf E|^2
\end{equation}
The electron fluid dynamics obey the continuity and momentum equations,

\begin{equation}
\partial_t n + \nabla\!\cdot(n\mathbf v)=0
\end{equation}

\begin{equation}
m_e(\partial_t + \mathbf v\!\cdot\!\nabla)\mathbf v = -e\mathbf E_{\rm low} - m_e\nu\mathbf v - \nabla U_p - \frac{1}{n}\nabla P
\end{equation}
where \(\mathbf E_{\rm low}\) is the slowly varying electric field induced by charge separation and \(P\) is the electron pressure.

Assuming small perturbations, we linearize the equations around equilibrium and introduce the electron displacement \(\boldsymbol{\xi}\) defined by \(\tilde{\mathbf v} = \partial_t \boldsymbol{\xi}\). Conservation of particles gives

\begin{equation}
\tilde n = -n_0 \nabla\!\cdot\boldsymbol{\xi}
\end{equation}

The linearized momentum equation then becomes
\begin{equation}
m_e \partial_t^2\boldsymbol{\xi} + m_e\nu \partial_t\boldsymbol{\xi}
= -e\mathbf E_{\rm low} - \nabla U_p - \frac{1}{n_0}\nabla P^{(1)}
\end{equation}
Taking the divergence and substituting Poisson’s law,

\begin{equation}
\nabla\!\cdot\mathbf E_{\rm low} = -\frac{e}{\varepsilon_0}\tilde n = \frac{e n_0}{\varepsilon_0}\nabla\!\cdot\boldsymbol{\xi}
\end{equation}
we obtain a driven oscillator equation for the density perturbation:

\begin{equation}
\partial_t^2 \tilde n + \nu \partial_t \tilde n + \omega_p^2 \tilde n
= \frac{n_0}{m_e}\nabla^2 U_p + \frac{1}{m_e}\nabla^2 P^{(1)}, \qquad
\omega_p^2 = \frac{e^2 n_0}{\varepsilon_0 m_e}
\end{equation}
For a cold plasma (\(P^{(1)} \approx 0\)), this simplifies to

\begin{equation}
\partial_t^2 \tilde n + \nu \partial_t \tilde n + \omega_p^2 \tilde n
= \frac{n_0}{m_e}\nabla^2 U_p
\end{equation}

In the quasi-static limit where the envelope evolves slowly compared with \(\omega_p\), the inertial term can be neglected, giving

\begin{equation}
\tilde n(\mathbf r,t)\;\approx\;\frac{n_0(\mathbf r)}{m_e\omega_p^2(\mathbf r)}\,\nabla^2 U_p(\mathbf r,t)
= \frac{\varepsilon_0}{4m_e\omega_0^2}\,\nabla^2|\mathbf E|^2
\end{equation}
Including finite-temperature effects introduces a pressure term \(P^{(1)}=k_B T_e \tilde n\), leading to a modified response in Fourier space:

\begin{equation}
\left(-\omega^2 + i\nu\omega + \omega_p^2 + \frac{\gamma k^2 k_B T_e}{m_e}\right)\tilde n(\mathbf k,\omega)
= -\frac{n_0 k^2}{m_e}U_p(\mathbf k,\omega)
\end{equation}
Hence,

\begin{equation}
\tilde n(\mathbf k,\omega)
= -\frac{n_0 k^2/m_e}{\omega_p^2 - \omega^2 + i\nu\omega + 3k^2v_{\rm th}^2}U_p(\mathbf k,\omega)
\end{equation}
where \(v_{\rm th}=\sqrt{k_B T_e/m_e}\).  
Finally, projecting the perturbation onto plasma eigenmodes gives

\begin{equation}
\begin{aligned}
\tilde n_m(z,t) 
&= \int \phi_m^*(\mathbf r)\, \tilde n(\mathbf r,z,t)\, d^2r, \\[1mm]
&\approx \int \phi_m^*(\mathbf r)\, \nabla^2\!\Big(\frac{e^2}{4 m_e \omega_0^2} |\Psi|^2 \Big) d^2r
\end{aligned}
\end{equation}
provides the nonlinear coupling coefficients entering the operator-based laser equation. In this form, the ponderomotive nonlinearity manifests as a quadratic self-interaction term in the field envelope, linking plasma dynamics to the evolution of the optical mode amplitudes.

\section{Plasma Oscillation Operator}

The plasma oscillation operator $\hat{\Omega}_p^2$ provides a compact representation of the collective electron response to small perturbations around equilibrium. Starting from the linearized cold-fluid equations for an electron plasma with equilibrium density $n_0(\mathbf{r})$, and assuming immobile ions, one obtains:
\begin{align}
\partial_t \tilde{n} + \nabla \!\cdot (n_0 \tilde{\mathbf{v}}) &= 0, \nonumber \\
m_e \partial_t \tilde{\mathbf{v}} &= - e \mathbf{E}_{\mathrm{low}}, \nonumber \\
\nabla \!\cdot \mathbf{E}_{\mathrm{low}} &= - \frac{e}{\varepsilon_0} \tilde{n}.
\end{align}
Eliminating $\tilde{\mathbf{v}}$ and $\mathbf{E}_{\mathrm{low}}$ yields a second-order equation for the density perturbation:
\begin{equation}
\partial_t^2 \tilde{n}(\mathbf{r},t) + \mathcal{L}_{\mathrm{plasma}} \tilde{n}(\mathbf{r},t) = 0,
\end{equation}
where $\mathcal{L}_{\mathrm{plasma}}$ is a self-adjoint operator determined by the plasma geometry and equilibrium density. Its eigenfunctions $\phi_m(\mathbf{r})$ and eigenvalues $\omega_{p,m}^2$ satisfy
\begin{equation}
\mathcal{L}_{\mathrm{plasma}} \phi_m = \omega_{p,m}^2 \phi_m,
\end{equation}
defining the natural plasma modes and their characteristic oscillation frequencies. The general solution can then be expanded in this orthonormal basis:
\begin{equation}
\tilde{n}(\mathbf{r},t) = \sum_m \delta n_m(t) \phi_m(\mathbf{r}),
\end{equation}
so that each modal amplitude evolves as a simple harmonic oscillator:
\begin{equation}
\ddot{\delta n}_m + \omega_{p,m}^2 \delta n_m = 0.
\end{equation}
The operator $\hat{\Omega}_p^2$ is therefore constructed as
\begin{equation}
\hat{\Omega}_p^2 = \sum_m \omega_{p,m}^2 \ket{\phi_m}\bra{\phi_m},
\end{equation}
which acts on any density perturbation through
\begin{equation}
\hat{\Omega}_p^2 \ket{\delta n} = \sum_m \omega_{p,m}^2 \delta n_m \ket{\phi_m}.
\end{equation}
This formulation generalizes the scalar plasma frequency $\omega_p^2 = e^2 n_0 / \varepsilon_0 m_e$ to arbitrary geometries and nonuniform plasmas, including boundary conditions and mode coupling. In the presence of collisions or external driving forces, each mode obeys:
\begin{equation}
\ddot{\delta n}_m + \nu_m \dot{\delta n}_m + \omega_{p,m}^2 \delta n_m = S_m(t),
\end{equation}
where $\nu_m$ is the damping rate and $S_m(t)$ is the projection of the driving term, such as the ponderomotive source, onto mode $m$. The thermal correction modifies the eigenvalue as $\omega_{p,m}^2 \to \omega_{p,m}^2 + k_m^2 v_{\mathrm{th}}^2$, while kinetic effects, including Landau damping, introduce an imaginary component $\omega_{p,m} \rightarrow \omega_{p,m} - i \gamma_{L,m}$.

In practice, $\mathcal{L}_{\mathrm{plasma}}$ can be discretized using finite difference, finite element, or spectral methods. The resulting matrix eigenproblem $\mathbf{D}\mathbf{u}_m = \omega_{p,m}^2 \mathbf{u}_m$ provides discrete modes and frequencies, from which the operator can be reconstructed numerically. Once normalized, the modes form a complete orthogonal basis, allowing reconstruction of any perturbation as $\tilde{n}(\mathbf{r},t) = \sum_m \delta n_m(t) \phi_m(\mathbf{r})$.

The operator $\hat{\Omega}_p^2$ thus serves as the fundamental link between the microscopic plasma equations and the macroscopic description of plasma oscillations. It provides a structured approach to describe wave excitation, damping, and mode coupling in both uniform and nonuniform plasmas, and forms the basis for the self-consistent treatment of laser–plasma interaction in operator form.

\section{From Maxwell–Plasma Dynamics to Operator Approach }

In this section, we demonstrate that the operator-based formalism introduced for LPWA is equivalent to the original coupled Maxwell--plasma partial differential equations (PDEs) under a mode decomposition. 

The dynamics of the electromagnetic field and plasma response are governed by \cite{sprangle1990nonlinear}
\begin{align}
\nabla^2 \mathbf{E} - \frac{1}{c^2}\frac{\partial^2 \mathbf{E}}{\partial t^2} &= \mu_0 \frac{\partial^2 \mathbf{P}}{\partial t^2}, \\
\frac{\partial^2 \delta n}{\partial t^2} + \omega_p^2 \delta n &= - \frac{e n_0}{m_e} \nabla \cdot |\mathbf{E}|^2,
\end{align}
where $\mathbf{E}$ is the electric field, $\mathbf{P}$ is the plasma polarization, $\delta n$ is the plasma density perturbation, and $\omega_p$ is the plasma frequency.

Under the paraxial approximation, the electromagnetic field is predominantly transverse and can be expanded in a set of orthonormal transverse modes $\{\psi_n(\mathbf{r}_\perp)\}$:
\begin{equation}
\mathbf{E}(\mathbf{r}_\perp,z,t) = \sum_n A_n(z) \psi_n(\mathbf{r}_\perp) e^{i(\beta_n z - \omega_0 t)},
\end{equation}
where $A_n(z)$ are the mode amplitudes and $\beta_n$ are the propagation constants. Similarly, the plasma density perturbation can be expanded as
\begin{equation}
\delta n(\mathbf{r}_\perp,z,t) = \sum_m \delta n_m(z) \phi_m(\mathbf{r}_\perp) e^{-i\omega_0 t},
\end{equation}
with $\{\phi_m\}$ forming an orthonormal basis for the plasma response.

Substituting these expansions into the Maxwell--plasma PDEs and projecting onto the transverse modes yields coupled ordinary differential equations (ODEs) for the mode amplitudes:
\begin{align}
i \frac{d A_n}{dz} &= \sum_m K_{nm} A_m - \frac{\omega_0^2}{2 k_0 c^2} \sum_m N_{nm}[A] A_m, \\
\ddot{\delta n_m} + \omega_{p,m}^2 \delta n_m &= - \sum_n \alpha_{mn} |A_n|^2,
\end{align}
where the coupling coefficients are defined as
\begin{equation}
\begin{aligned}
K_{nm} &= \langle \psi_n | \hat{K} | \psi_m \rangle, &
N_{nm}[A] &= \langle \psi_n | \hat{N}[A] | \psi_m \rangle, \\[1mm]
\alpha_{mn} &= \langle \phi_m | \hat{\alpha} | \psi_n \rangle.
\end{aligned}
\end{equation}
Defining abstract state vectors in the mode space,
\begin{equation}
\ket{\Psi} = \sum_n A_n \ket{\psi_n}, \quad 
\ket{\delta n} = \sum_m \delta n_m \ket{\phi_m},
\end{equation}
and introducing the operators
\begin{equation}
\hat{K}, \quad \hat{N}[\Psi], \quad \hat{\Omega}_p^2, \quad \hat{\alpha},
\end{equation}
the evolution equations can be compactly written as
\begin{align}
i \partial_z \ket{\Psi} &= \hat K \ket{\Psi} - \frac{\omega_0^2}{2 k_0 c^2} \hat N[\Psi] \ket{\Psi}, \\
\ddot{\ket{\delta n}} + \hat \Omega_p^2 \ket{\delta n} &= - \hat \alpha |\Psi|^2.
\end{align}
Expanding these operator equations back in the mode basis reproduces exactly the ODEs derived from the PDEs, thus providing a mathematical validation of the operator formalism. This equivalence ensures that linear perturbations, nonlinear feedback, and plasma oscillations are fully described within the operator framework.

\section{Operator-Based Full-Vector Formalism for LPWA}

In LPWA, the longitudinal electric field component \(E_z\) plays a crucial role in wake formation and particle acceleration. To account for this, the electromagnetic field must be expanded in a vector mode basis rather than purely transverse modes. Each mode satisfies Maxwell’s equations in the plasma environment and includes both transverse and longitudinal components.

The electromagnetic field can be represented as
\begin{equation}
\mathbf{E}(\mathbf{r}, z) = \sum_n A_n(z)\, \boldsymbol{\psi}_n(\mathbf{r})\, e^{i \beta_n z},
\end{equation}
where each mode function is a vector
\begin{equation}
\boldsymbol{\psi}_n(\mathbf{r}) =
\begin{pmatrix}
\psi_{n,x}(\mathbf{r}) \\
\psi_{n,y}(\mathbf{r}) \\
\psi_{n,z}(\mathbf{r})
\end{pmatrix}.
\end{equation}
Each mode satisfies the generalized eigenvalue equation derived from the curl–curl form of Maxwell’s equations:
\begin{equation}
\nabla \times (\nabla \times \boldsymbol{\psi}_n) = k_0^2 \epsilon(\mathbf{r}) \boldsymbol{\psi}_n = \beta_n^2 \boldsymbol{\psi}_n,
\end{equation}
together with the transversality condition
\begin{equation}
\nabla \cdot [\epsilon(\mathbf{r}) \boldsymbol{\psi}_n] = 0.
\end{equation}
To compactly describe the coupled dynamics, we define a vector field state:
\begin{equation}
\ket{\Psi} =
\begin{pmatrix}
\ket{\Psi_T} \\
\ket{\Psi_L}
\end{pmatrix},
\end{equation}
where \(\ket{\Psi_T}\) represents the transverse (x,y) components and \(\ket{\Psi_L}\) the longitudinal component. The evolution along the propagation axis \(z\) is then expressed as
\begin{equation}
i \partial_z
\begin{pmatrix}
\ket{\Psi_T} \\
\ket{\Psi_L}
\end{pmatrix}
=
\begin{pmatrix}
\hat{K}_{TT} & \hat{K}_{TL} \\
\hat{K}_{LT} & \hat{K}_{LL}
\end{pmatrix}
\begin{pmatrix}
\ket{\Psi_T} \\
\ket{\Psi_L}
\end{pmatrix}
- \frac{\omega_0^2}{2 k_0 c^2}
\begin{pmatrix}
\hat{N}_T[\Psi] \\
\hat{N}_L[\Psi]
\end{pmatrix}.
\end{equation}
Here, \(\hat{K}_{TT}\) describes transverse propagation and diffraction, \(\hat{K}_{LL}\) includes longitudinal propagation corrections, and \(\hat{K}_{TL}, \hat{K}_{LT}\) account for mode coupling between transverse and longitudinal fields. The nonlinear operators \(\hat{N}_T[\Psi], \hat{N}_L[\Psi]\) describe ponderomotive feedback and charge redistribution within the plasma.

The plasma density perturbation is similarly expanded in the plasma eigenmode basis \(\{ \phi_m \}\):
\begin{equation}
\ket{\delta n} = \sum_m \delta n_m(z) \ket{\phi_m},
\end{equation}
and the ponderomotive source term couples both field components:
\begin{align}
\ddot{\ket{\delta n}} + \hat{\Omega}_p^2 \ket{\delta n} &=
- \hat{\alpha}_T |\Psi_T|^2 
- \hat{\alpha}_L |\Psi_L|^2 \nonumber \\
&\quad - \hat{\alpha}_{TL} (\Psi_T^* \Psi_L + \Psi_L^* \Psi_T),
\end{align}
where \(\hat{\alpha}_{TL}\) represents the cross-coupling between transverse and longitudinal fields that drives wake formation.

By expanding in the eigenbasis \(\{ \boldsymbol{\psi}_n \}\), the system reduces to a set of scalar coupled equations for each mode amplitude:
\begin{equation}
i \frac{d A_n}{dz} = \sum_m K_{nm} A_m - \frac{\omega_0^2}{2 k_0 c^2} \sum_m N_{nm}[A] A_m,
\end{equation}
with
\begin{equation}
K_{nm} = \langle \boldsymbol{\psi}_n | \hat{K} | \boldsymbol{\psi}_m \rangle, \quad
N_{nm}[A] = \langle \boldsymbol{\psi}_n | \hat{N}[\Psi] | \boldsymbol{\psi}_m \rangle.
\end{equation}
The longitudinal–transverse coupling is manifested in the off-diagonal elements \(K_{n_L, n_T}\).

Th full-vector formalism unifies transverse and longitudinal dynamics in a compact operator framework, offering both analytical insight and numerical efficiency. Each physical mechanism appears as mode–mode interactions, facilitating systematic inclusion of nonlinearities and perturbative effects. The inclusion of longitudinal modes within this operator-based framework provides a mathematically complete and physically transparent description of LPWA.

\section{Applying Bloch–Floquet Theory to Operator Formalism}

LPWA often produces periodic structures in the plasma due to the oscillatory wake driven by a high-intensity laser \cite{sprangle1988laser}. These periodic modulations of the plasma density and refractive index can be analyzed using concepts from Bloch-Floquet theory, widely used in solid-state physics for electrons in periodic potentials and in photonics for periodic dielectric media\cite{joannopoulos2008molding}, \cite{russell2003photonic}.

Consider a plasma wake with a periodic longitudinal structure along the propagation axis $z$, with period $\Lambda$, often close to the plasma wavelength $\lambda_p$. The plasma density can be decomposed as \cite{sprangle1988laser}
\begin{equation}
\begin{aligned}
n_0(\mathbf r_\perp, z) &= n_\mathrm{bkg}(\mathbf r_\perp) + \delta n(\mathbf r_\perp, z), \\[1mm]
\delta n(\mathbf r_\perp, z+\Lambda) &= \delta n(\mathbf r_\perp, z).
\end{aligned}
\end{equation}
where $\mathbf r_\perp=(x,y)$ are transverse coordinates. The corresponding modulation of the refractive index or permittivity is
\begin{equation}
\epsilon(\mathbf r_\perp, z) = \epsilon_0(\mathbf r_\perp) + \Delta \epsilon(\mathbf r_\perp, z), \
\Delta \epsilon(\mathbf r_\perp, z+\Lambda) = \Delta \epsilon(\mathbf r_\perp, z),
\end{equation}
which can be expanded in a Fourier series along $z$ \cite{russell2003photonic}:
\begin{equation}
\Delta \epsilon(\mathbf r_\perp, z) = \sum_G \widetilde{\Delta \epsilon}_G(\mathbf r_\perp) e^{i G z}, \quad G = \frac{2 \pi p}{\Lambda}, \, p \in \mathbb Z.
\end{equation}
Similarly, the density modulation can be expressed through a Fourier decomposition
\begin{equation}
\delta n(\mathbf r_\perp, z) = \sum_G \widetilde{n}_G(\mathbf r_\perp) e^{i G z}.
\end{equation}

For the electromagnetic-wave analog, the curl-curl Maxwell eigenproblem in a periodic dielectric \cite{joannopoulos2008molding},
\begin{equation}
\nabla \times \big(\nabla \times \mathbf E(\mathbf r) \big) = \frac{\omega^2}{c^2} \epsilon(\mathbf r) \mathbf E(\mathbf r), \quad \epsilon(z+\Lambda) = \epsilon(z),
\end{equation}
supports Bloch solutions
\begin{equation}
\mathbf E_{\mathbf k}(\mathbf r) = e^{i \mathbf k \cdot \mathbf r} \mathbf u_{\mathbf k}(\mathbf r), \quad 
\mathbf u_{\mathbf k}(\mathbf r+\Lambda \hat z) = \mathbf u_{\mathbf k}(\mathbf r),
\end{equation}
analogous to the periodic electron problem.
For operator-based LPWA formalism, the evolution equation along $z$ is
\begin{equation}
i \frac{\partial}{\partial z} \ket{\Psi(z)} = \hat{\mathcal L}(z) \ket{\Psi(z)}, \quad \hat{\mathcal L}(z+\Lambda) = \hat{\mathcal L}(z),
\end{equation}
where $\ket{\Psi(z)}$ is the laser envelope vector in the mode basis. Floquet's theorem guarantees solutions of the form \cite{joannopoulos2008molding}
\begin{equation}
\ket{\Psi(z)} = e^{i \mu z} \ket{u_\mu(z)}, \quad \ket{u_\mu(z+\Lambda)} = \ket{u_\mu(z)},
\end{equation}
with $\mu$ the Floquet exponent. The monodromy (transfer) matrix over one period is defined as
\begin{equation}
\mathcal M \equiv \mathcal U(\Lambda, 0), \quad \mathcal U(z_2,z_1) = \text{propagator of } i\partial_z \ket{\Psi} = \hat{\mathcal L}(z)\ket{\Psi},
\end{equation}
with eigenvalues $\lambda_j = e^{i \mu_j \Lambda}$.

For a single-harmonic longitudinal modulation $\Delta n(z) = \Delta n_0 \cos(G_0 z)$, the envelope is expanded as
\begin{equation}
\mathcal E(z) = A_+(z) e^{i k_0 z} + A_-(z) e^{-i k_0 z},
\end{equation}
leading to coupled-mode equations
\begin{equation}
\frac{d A_+}{dz} = i \delta A_+ + i \kappa A_-, \quad
\frac{d A_-}{dz} = - i \delta A_- + i \kappa^* A_+,
\end{equation}
with dispersion relation
\begin{equation}
\mu^2 = \delta^2 + |\kappa|^2, \quad \Delta \mu = 2 |\kappa|.
\end{equation}
For a general low-amplitude periodic potential $V(z) = \sum_G V_G e^{i G z}$ near degeneracy, the effective Hamiltonian is \cite{joannopoulos2008molding}, \cite{russell2003photonic}
\begin{equation}
H_\text{eff} = \begin{pmatrix} E_0(k) & V_{-G} \\ V_G & E_0(k+G) \end{pmatrix}, \quad
E_\pm = E_0 \pm |V_G|, \quad \Delta E = 2 |V_G|.
\end{equation}
This effective Hamiltonian describes the essential mode coupling induced by the periodic plasma modulation and provides a bridge to the operator-based description of the laser–plasma system.
The laser–plasma dynamics with periodic longitudinal modulation is expressed as
\begin{align}
i \frac{\partial}{\partial z} \ket{\Psi} &= \hat K(z) \ket{\Psi} - \frac{\omega_0^2}{2 k_0 c^2} \hat N[\Psi] \ket{\Psi}, \\
\ddot{\ket{\delta n}} + \hat \Omega_p^2(z) \ket{\delta n} &= - \hat \alpha |\Psi|^2,
\end{align}
with
\begin{equation}
\hat K(z+\Lambda) = \hat K(z), \quad \hat \Omega_p^2(z+\Lambda) = \hat \Omega_p^2(z),
\end{equation}
and mode-coupling coefficients
\begin{equation}
\kappa_{nm}^{(G)} = - \frac{\omega_0^2}{2 k_0 c^2} \int \psi_n^*(\mathbf r_\perp) \widetilde{\Delta \epsilon}_G(\mathbf r_\perp) \psi_m(\mathbf r_\perp) d^2 r_\perp.
\end{equation}

In a plasma with periodic longitudinal modulation, the laser naturally interacts with the plasma in a way that couples different optical modes. 
The effective Hamiltonian offers a simple way to describe these mode couplings near degeneracy, illustrating how energy is exchanged between modes and giving rise to phenomena such as band gaps. 
However, the laser--plasma system is inherently continuous along the propagation axis, so a more complete description requires the operator-based formalism. 
This formalism governs the full evolution of the laser envelope $\ket{\Psi(z)}$ under a periodic operator $\hat K(z)$, which automatically accounts for all mode couplings and the plasma's structure.

\section{Connection Between Operator Formalism and Invariant Subspaces}

The operator formalism can be derived from the structure of operator theory and functional analysis. In both contexts, the system dynamics can be represented by bounded linear operators acting on a Hilbert space \cite{enflo2023invariant},\cite{behtouei2023invariant} and the evolution of the state vector is governed by the structure of these operators and their invariant subspaces.

In the LPWA model, the laser field and plasma response are expanded over a set of modes \ref{laser-field},\ref{plasma-field} which evolve under the coupled operator equations \ref{plasma} and \ref{laser}. The operators form a dynamical system on a Hilbert space $\mathcal{H}$ of field and plasma modes.
A key point is that in LPWA, the laser and plasma dynamics can be represented in terms of modal components, with linear operators defining invariant subspaces where modes would evolve independently. Nonlinear interactions, however, break this invariance, coupling the modes and producing energy transfer, instabilities, or complex spatial patterns. This highlights that the problem of finding invariant subspaces in the LPWA operator formalism is equivalent to identifying stable or quasi-stable sets of modes in the plasma-structures that  remain approximately independent despite nonlinear effects.

This observation connects the LPWA operator formalism with the Invariant Subspace Problem in operator theory. Invariant subspaces correspond physically to sets of modes that evolve independently under the propagation operator. If $\mathcal{S} \subset \mathcal{H}$ satisfies $\hat{\mathcal{H}}\mathcal{S} \subseteq \mathcal{S}$, then $\mathcal{S}$ is an invariant mode manifold, a condition often broken in nonlinear plasma dynamics, where energy transfer among modes occurs. The search for invariant subspaces thus corresponds to identifying stable or quasi-stable modal structures in laser–plasma interaction.

The evolution of the laser field envelope in operator form can be equivalently written as
\begin{equation}
i\,\frac{\partial}{\partial z}\ket{\Psi} = 
\hat{\mathcal{H}}[\Psi]\ket{\Psi}, \qquad
\hat{\mathcal{H}}[\Psi] = 
\hat K - \frac{\omega_0^2}{2 k_0 c^2}\,\hat N[\Psi].
\end{equation}
The total Hilbert space $\mathcal{H}=L^2(\mathbb{R}^2)$ contains all admissible transverse field distributions, and each eigenmode of $\hat K$ corresponds to a one-dimensional invariant subspace:
\begin{equation}
\hat K \ket{\psi_n} = \beta_n \ket{\psi_n}, \qquad
\hat K\big(\mathrm{span}\{\ket{\psi_n}\}\big)\subseteq \mathrm{span}\{\ket{\psi_n}\}.
\end{equation}
These subspaces describe pure mode propagation in the absence of plasma coupling. When the nonlinear operator $\hat N[\Psi]$ becomes active, it mixes these invariant subspaces, leading to energy transfer and mode coupling. The field no longer evolves within a single mode subspace but explores a larger manifold within $\mathcal{H}$, governed by the nonlinear operator structure:
\begin{equation}
\hat N[\Psi]\ket{\psi_n} = \sum_m N_{mn}[\Psi]\,\ket{\psi_m}.
\end{equation}
This corresponds to the breakdown of stable modal propagation as the plasma modifies the refractive index, leading to mode beating, wake excitation, and nonlinear self-focusing. 
Provided that a nontrivial invariant subspace $\mathcal{M}\subset\mathcal{H}$ exists such that 
$\hat{\mathcal{H}}[\Psi]\mathcal{M}\subseteq\mathcal{M}$, 
then $\mathcal{M}$ represents a dynamically stable configuration of coupled laser--plasma states, 
potentially corresponding to stationary wake structures or self-guided modes. 
The absence of such invariant subspaces implies full intermodal mixing and a transition toward complex or chaotic plasma--field evolution. 
The mathematical properties of $\hat{\mathcal{H}}[\Psi]$, such as self-adjointness, boundedness, and spectral completeness, 
are therefore crucial in determining whether stable subspaces exist. 
When no invariant subspaces persist, the system can exhibit hypercyclic behavior, where even small differences in the initial laser field lead to dramatically different evolutions.

A bounded operator $\hat{T}$ on $\mathcal{H}$ is called hypercyclic if there exists a vector $\ket{\Psi_0}$ 
whose orbit $\{\hat{T}^n\ket{\Psi_0} : n\in\mathbb{N}\}$ is dense in $\mathcal{H}$.\cite{behtouei2023invariant2}
Physically, this condition corresponds to a regime where infinitesimal changes in the initial laser field 
lead to completely different evolutions, analogous to chaos in classical dynamical systems. 
In LPWA, such hypercyclic-like dynamics can arise when nonlinear coupling between modes 
drives stochastic energy exchange between the laser and the plasma, 
leading to turbulence or self-organized complexity in the wakefield evolution.

\section{Integration of Operator Formalism with Artificial Intelligence}

The operator formalism developed in the previous section provides a mathematically compact and physically transparent framework for describing the coupled evolution of the laser field and the plasma response. This formulation is also ideally suited for integration with modern artificial intelligence (AI) methods, particularly neural operator architectures, which can efficiently learn and reproduce mappings between functional spaces governed by complex nonlinear dynamics.

From a computational standpoint, the most challenging components of the coupled evolution of the laser envelope and the plasma density perturbation, \ref{plasma} and \ref{laser}, are the nonlinear and nonlocal operators $\hat N[\Psi]$ and $\hat \alpha$. These terms depend on the full spatiotemporal field structure, local plasma response, and nonlocal averaging effects due to the ponderomotive force. Traditional numerical solvers for these operators often require high-resolution meshes and iterative coupling between Maxwell and fluid (or kinetic) equations, leading to prohibitive computational cost for realistic 3D geometries.

Neural operator models offer a powerful alternative. A neural operator $\mathcal{G}_\theta$ is a trainable functional mapping \cite{kovachki2023neural},\cite{kovachki2021universal}
\[
\mathcal{G}_\theta : \Psi(\mathbf r,z) \mapsto \hat N[\Psi](\mathbf r,z),
\]
parametrized by learnable weights $\theta$. Once trained, $\mathcal{G}_\theta$ approximates the nonlinear operator $\hat N$ with high fidelity, enabling ultra-fast evaluation during simulation or control. Similarly, a second neural operator $\mathcal{A}_\phi$ can learn the mapping
\[
\mathcal{A}_\phi : |\Psi|^2(\mathbf r,z) \mapsto \hat \alpha |\Psi|^2,
\]
effectively reproducing the plasma source term driving wake formation. These models can be trained using data from high-fidelity particle-in-cell (PIC) or fluid simulations, where the true operator actions are known for various laser and plasma configurations.

In practice, these neural operators can be embedded within the operator equations as
\begin{align}
i \frac{\partial \ket{\Psi}}{\partial z} &= 
\hat K \ket{\Psi} - \frac{\omega_0^2}{2 k_0 c^2}\, \mathcal{G}_\theta[\Psi] \ket{\Psi}, \\
\ddot{\ket{\delta n}} + \hat \Omega_p^2 \ket{\delta n} &= 
- \mathcal{A}_\phi[|\Psi|^2],
\end{align}
where $\mathcal{G}_\theta$ and $\mathcal{A}_\phi$ replace the explicit evaluation of $\hat N$ and $\hat \alpha$ respectively.  
This substitution allows for efficient reduced-order modeling that preserves the underlying operator structure and physical interpretability of the formalism. By approximating $\hat N$ and $\hat \alpha$ with neural operators, one effectively teaches an AI system to “learn” the plasma’s nonlinear memory, the nonlocal and time-delayed response that governs wake growth, mode coupling, and saturation. Such neural operator models can also be made differentiable, allowing gradient-based optimization and control. For example, the AI-enhanced system can predict how variations in input pulse shape or capillary geometry affect wake amplitude, energy transfer efficiency, and beam quality. 

The combination of physics-based operator structure and data-driven approximation defines a hybrid model
\[
\hat{\mathcal{H}}_{\text{hyb}} =
\hat K - \frac{\omega_0^2}{2 k_0 c^2}\, \mathcal{G}_\theta[\Psi],
\]
which preserves the Hermitian and symmetry properties of the underlying system while replacing computationally expensive nonlinear evaluations with trained models. The hybrid operator $\hat{\mathcal{H}}_{\text{hyb}}$ acts as an effective Hamiltonian for the coupled laser–plasma system, enabling rapid and accurate prediction of field evolution under varying conditions. In this sense, neural operators do not replace the physics, they refine and accelerate it. The formalism provides a mathematically grounded interface between analytical theory and data-driven modeling, where each operator retains its physical meaning while its functional form is learned from empirical data.

Integrating operator formalism with neural operator AI architectures thus enables scalable, reduced-order modeling of LPWA. It allows prediction and control of nonlinear, multidimensional laser–plasma interactions at a fraction of the computational cost of traditional solvers. The resulting framework forms the foundation for intelligent, physics-informed accelerator design, where operator-based mathematics and artificial intelligence work in synergy to describe and control high-field plasma dynamics in next-generation accelerators.
\section{Conclusion}

In this work, we have developed a comprehensive operator-based framework for describing laser-plasma wakefield acceleration, bridging the gap between traditional Maxwell–plasma PDEs and a compact, mode-centric representation. By decomposing both the laser field and the plasma density perturbations into orthonormal modal bases, we have shown that the coupled dynamics of linear imperfections, nonlinear plasma response, and ponderomotive driving can be expressed through a set of well-defined operators: $\hat K$ for transverse modal, $\hat N$ for nonlinear feedback, $\hat \Omega_p^2$ for intrinsic plasma oscillations, and $\hat \alpha$ for the ponderomotive source.

This formalism offers several advantages. It provides a transparent picture of energy transfer between modes, describes both transverse and longitudinal fields, and supports complex plasma geometries and guiding structures. By integrating techniques such as Bloch–Floquet theory, the operator approach also enables the systematic analysis of periodic plasma modulations and mode coupling effects, which are essential for understanding wakefield dynamics in realistic experimental scenarios.

Ultimately, the operator-based description not only reproduces the results of full PDE simulations but also enhances analytical insight, reduces computational complexity, and offers a versatile platform for exploring advanced LPWA configurations. This approach provides the way for optimized accelerator design, multi-mode control, and integration with AI-based modeling strategies, providing a robust foundation for future developments in compact, high-gradient plasma accelerators.

\section*{Acknowledgment}

The research leading to these results has received funding from: 
EuPRAXIA (Grant Agreement No. 101079773, European Union Horizon Europe INFRA research and innovation program) and 
PACRI (Grant Agreement No. 101188004, European Union Horizon Europe INFRA research and innovation program).

\section*{References}

\bibliographystyle{apsrev4-2}

\end{document}